\documentclass[fleqn,twoside]{article}
\usepackage{amsmath}
\usepackage{espcrc2}
\usepackage{graphicx}

\title{Relation between color-deconfinement and chiral restoration}
\author{Kenji Fukushima
\thanks{We acknoledge the support of the Japan Society for the
Promotion of Science for Young Scientists. The present address of the
author is: Center for Theoretical Physics, Massachusetts Institute of
Technology, Cambridge, MA.}
\address{Department of Physics, University of Tokyo,
         7-3-1 Hongo, Bunkyo-ku, Tokyo 113-0033, Japan}}

\begin{document}

\begin{abstract}
We discuss the relation between the Polyakov loop and the chiral order
parameter at finite temperature by using an effective model. We
clarify why and how the pseudo-critical temperature associated with
the Polyakov loop should coincide with that of the chiral condensate.
\vspace{1pc}
\end{abstract}

\maketitle


Quantum Chromodynamics (QCD) is believed to undergo phase transitions
from hadronic matter to a quark-gluon plasma (QGP) at high
temperature. The QGP phase transitions can consist of two distinct
transitions -- the deconfinement transition and the chiral phase
transition. Several theoretical studies \cite{cas79} have clarified
that confinement would be closely related to chiral dynamics and the
confined phase should be accompanied by spontaneous breaking of the
chiral symmetry. This means that the deconfinement transition
temperature must be lower than the chiral phase transition
temperature.

Actually, in the presence of massive fundamental quarks, neither the
deconfinement transition nor the chiral phase transition has any
well-defined order parameter. In the lattice QCD simulations
\cite{kar94}, it has been observed that the approximate order
parameters for two phase transitions, that is, the Polyakov loop
and the chiral condensate both show crossover behavior like the
magnetization under an external magnetic field. We cannot define the
critical temperature any more. Instead, we can only locate the
``critical point'' by the peak of susceptibility, namely the
pseudo-critical temperature. The lattice QCD simulations have
revealed that the respective peaks of the Polyakov loop and the chiral
susceptibilities are located at the same temperature. It has been a
puzzle why and how those pseudo-critical temperatures should be
identical.

A simple explanation has been proposed by Satz \cite{sat98}. At low
temperature with spontaneous chiral symmetry breaking, the center
symmetry breaking is suppressed due to constituent quark mass. The
Polyakov loop is small as a result. In the chiral symmetric phase at
high temperature, the Polyakov loop can become large. Therefore the
constituent quark mass should govern the Polyakov loop behavior rather
than deconfinement would. In fact, Gocksch and Ogilvie \cite{goc85}
had constructed an effective model based on the strong coupling
expansion and argued that Satz's argument is embodied
qualitatively. Recently it has been shown by the present author
\cite{fuk03} that the Gocksch-Ogilvie model works pretty well and
Satz's argument makes too much of the chiral phase transition. The
Polyakov loop dynamics is also important to understand the coincidence
of pseudo-critical temperatures.


The Gocksch-Ogilvie model is given by the effective action
\cite{goc85,fuk03}
\begin{align}
 & S_{\text{eff}}^{(r)}[L,\lambda] =-\mathrm{e}^{-\sigma a/T}
  \sum_{\text{n.n.}}\mathrm{Tr_c} L(\vec{n})\mathrm{Tr_c}
  L^\dagger(\vec{m}) \notag\\
 & +\frac{\text{dim}(r)}{2}
  \sum_{m,n}\lambda(n)V(n,m)\lambda(m) \notag\\
 & -\frac{N_{\text{f}}}{4}\sum_{\vec{n}}\mathrm{Tr_c}\ln\Bigl[
  \cosh(N_\tau E)+\frac{1}{2}\bigl(L^{(r)}+L^{(r)\dagger}\bigr)\Bigr],
\label{eq:action}
\end{align}
where $L(\vec{n})$ is the Polyakov loop in the fundamental
representation defined on the lattice by
\begin{equation}
 L(\vec{n})=\prod_{n_d=a}^{N_\tau a}U_d(\vec{n},n_d),
\end{equation}
in the $d$ dimensional space-time. $r$ is the representation of
dynamical quarks. The Polyakov loop in the adjoint representation
($r=\text{adj}$) can be expressed in terms of the fundamental one as
\begin{equation}
 L_{ab}^{\text{(adj)}}=2\mathrm{Tr_c}\bigl[t_a L t_b L^\dagger\bigr],
\end{equation}
where the matrices $t_a$ form a fundamental representation of the
$\mathrm{SU}(N_{\text{c}})$ group. $\lambda(n)$ is the meson field,
$\mathrm{Tr_c}$ the trace with respect to the color indices, and
$N_{\text{f}}$ the number of flavors which is fixed as
$N_{\text{f}}=2$ in the present analyses. $\text{dim}(r)$ is the
dimension of the $r$ representation, that is,
$\text{dim}(\text{fund})=N_{\text{c}}$ and
$\text{dim}(\text{adj})=N_{\text{c}}^2-1$. The meson hopping
propagator $V(n,m)$ and the quasi-quark energy (constituent quark
mass) $E$ are defined respectively as
\begin{align}
 & V(n,m)=\frac{1}{2(d-1)}\sum_{\hat{j}}\bigl(\delta_{n,m+\hat{j}}
  +\delta_{n,m-\hat{j}}\bigr), \notag\\
 & E=\sinh^{-1}\biggl(\sqrt{\frac{d-1}{2}}\lambda+m_q\biggr)
\end{align}
with $\hat{j}$ running from $1$ to $d-1$ only in the spatial
directions. $\sigma$ in Eq.~(\ref{eq:action}) is the string tension
fixed as $\sigma=(425\,\text{MeV})^2$. The lattice spacing $a$ and the
current quark mass $m_q$ are treated as the model parameters.


We shall adopt the mean field (Weiss) approximation for the Polyakov
loop $L$. In this approximation, we take account of the fluctuation of
the individual Polyakov loop surrounded by a constant mean field. As
for the chiral order parameter, $\lambda$ is simply treated as a
constant mean field. Detailed calculations are given in
Ref.~\cite{fuk03}.

\begin{table}[b]
\begin{center}
\begin{tabular}{c|cccc}
\hline \hline
 & $a^{-1}$ & $m_q$ & $T_{\text{d}}$ & $T_{\text{c}}$\\
\hline
 I & $432\,\text{MeV}$ & $5.7\,\text{MeV}$ & $208\,\text{MeV}$
  & $\sim187\,\text{MeV}$\\
\hline
 II & $333\,\text{MeV}$ & $7.4\,\text{MeV}$ & $270\,\text{MeV}$
  & $\sim230\,\text{MeV}$\\
\hline \hline
\end{tabular}
\end{center}
\caption{Model parameters and resulting outputs for fundamental
quarks.}
\label{tab:parameter}
\end{table}

In Table~\ref{tab:parameter}, we list the values of the model
parameters and several resulting outputs for fundamental
quarks. $T_{\text{d}}$ is the deconfinement temperature estimated from
the gluonic contribution alone and $T_{\text{c}}$ is the
pseudo-critical temperature defined by the peak of susceptibility.
Parameter I is determined to fit the pion mass and the $\rho$ meson
mass, while Parameter II reproduces the pion mass and the empirical
value of the deconfinement temperature, i.e.,
$T_{\text{d}}=270\,\text{MeV}$. These two parameter sets happen to
correspond to two typical cases, as explained below.

\begin{figure}[ht]
\vspace{-5mm}
\begin{center}
\includegraphics[width=3.6cm]{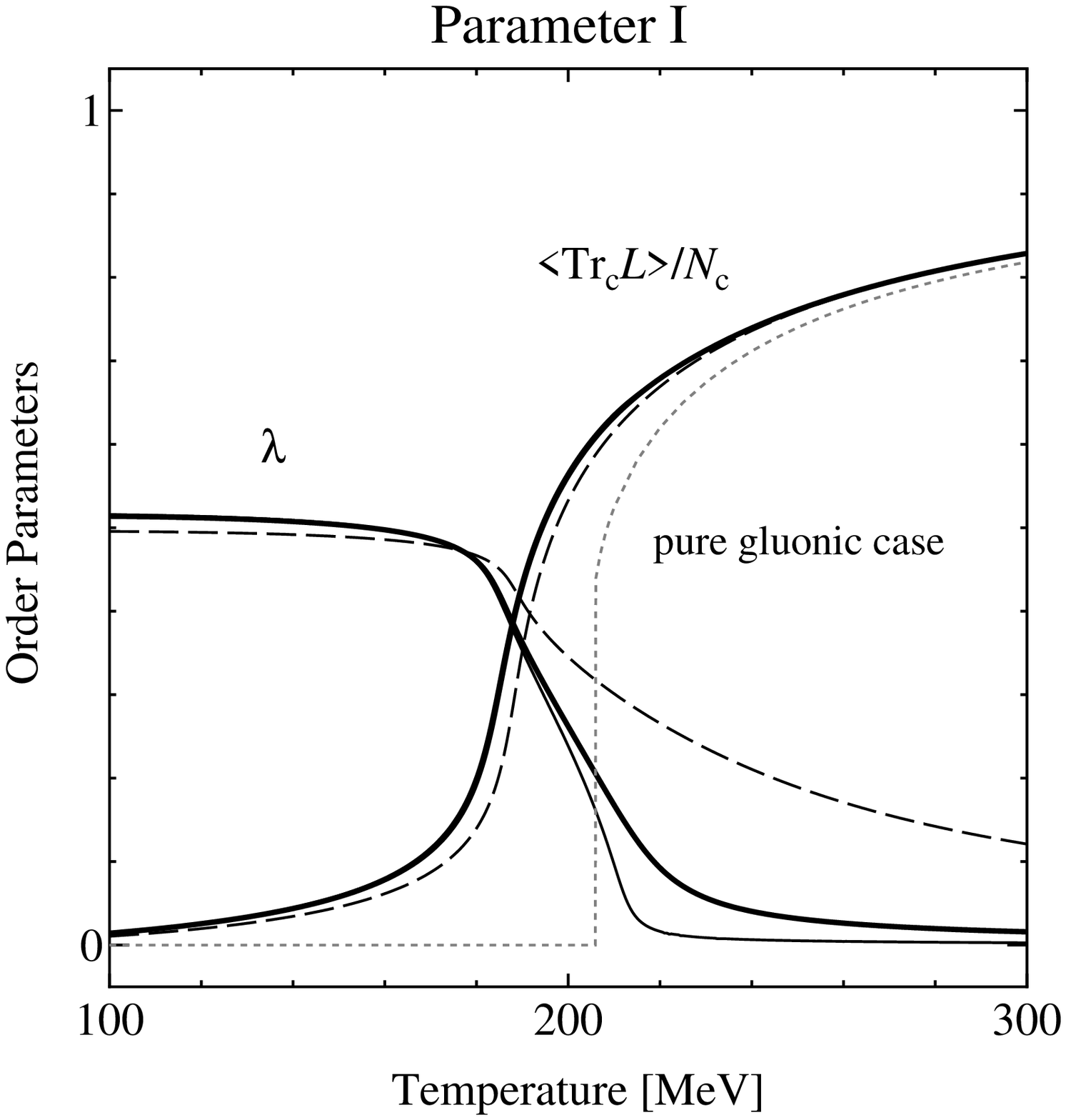}
\includegraphics[width=3.6cm]{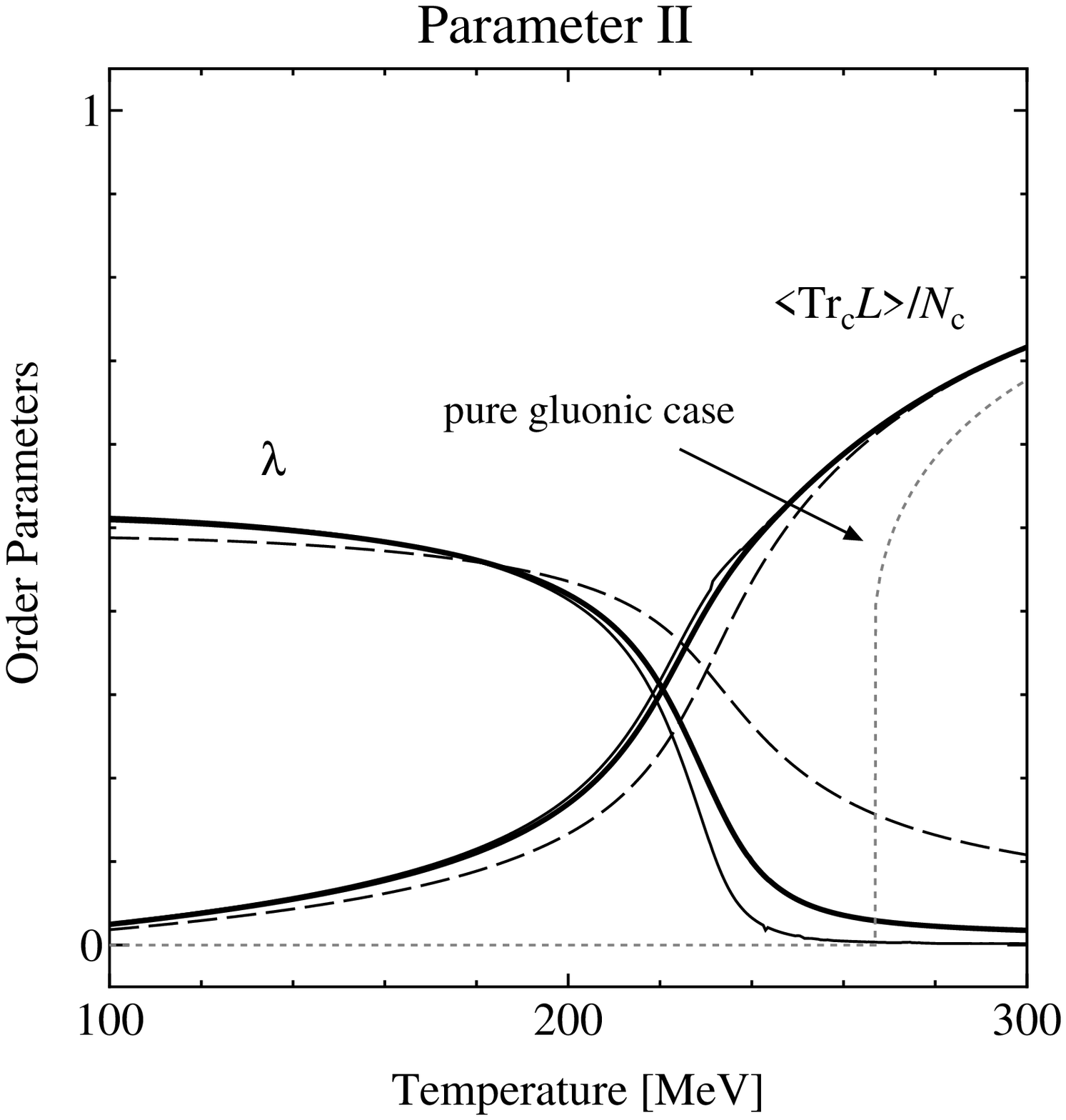}
\end{center}
\vspace{-10mm}
\caption{Order parameters: the thick solid curve for
$m_q=5.7\,\text{MeV}$ in Parameter I and $m_q=7.4\,\text{MeV}$ in
Parameter II, the thin solid and dashed curves for $m_q=1\,\text{MeV}$
and $50\,\text{MeV}$ respectively.}
\label{fig:order}
\vspace{-3mm}
\end{figure}

\begin{figure}[ht]
\begin{center}
\includegraphics[width=3.6cm]{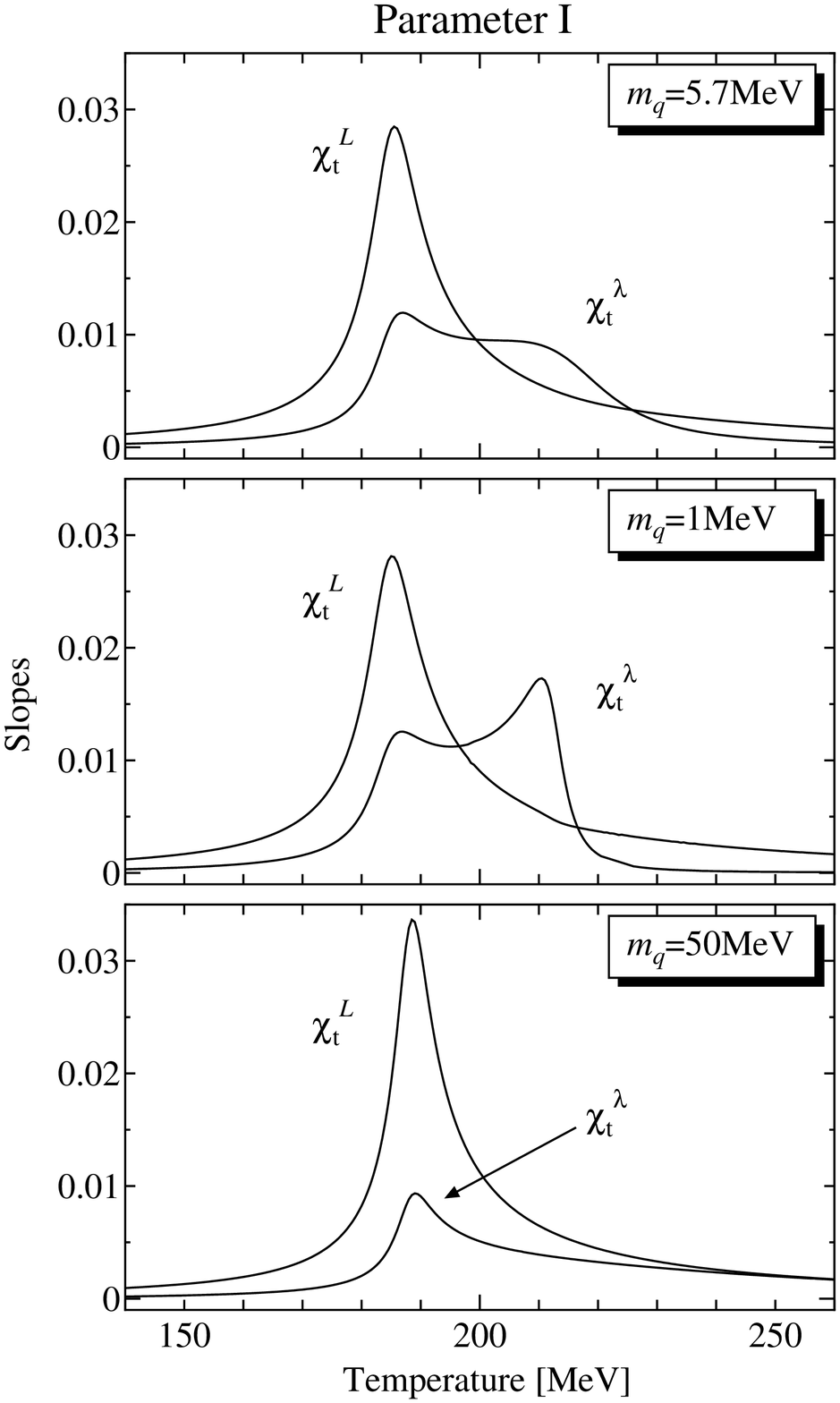}
\includegraphics[width=3.6cm]{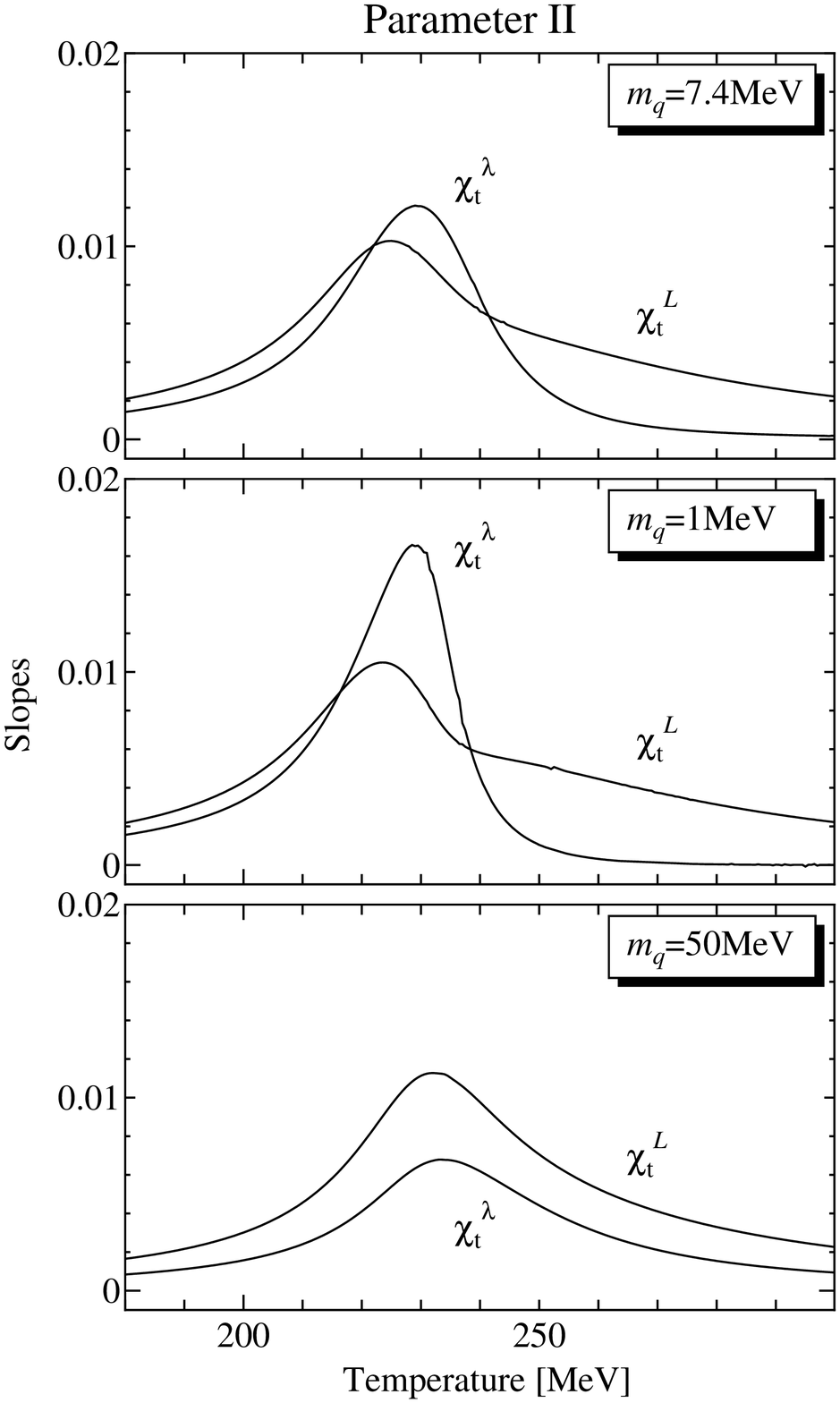}
\end{center}
\vspace{-10mm}
\caption{Temperature slopes}
\label{fig:slope}
\vspace{-3mm}
\end{figure}

The numerical results are shown in Fig.~\ref{fig:order} together with
the results for several different $m_q$. In both cases we can see the
order parameters for deconfinement and chiral restoration indicating
crossover behavior simultaneously. Although the results in
Fig.~\ref{fig:order} just seem to support Satz's argument, the
relation between color-deconfinement and chiral restoration has richer
physical contents, as inferred from Fig.~\ref{fig:slope}.

For $m_q=1\,\text{MeV}$ with Parameter I, two peaks appear clearly in
the behavior of slope,
$\chi_{\text{t}}^\lambda=-\partial\lambda/\partial T$. One peak around
$T_{\text{c}}\simeq 187\,\text{MeV}$ corresponds to the remnant of the
first order deconfinement transition. The other around $T\simeq
210\,\text{MeV}$ stems from the second order chiral phase transition,
which agrees with the thin curve in Fig.~\ref{fig:order}. Therefore
the coincidence of the pseudo-critical temperatures signifies not the
chiral phase transition but the deconfinement transition. In this
sense, we can say that the Deconfinement Dominance is realized in the
simultaneous crossovers.

The Deconfinement Dominance is seen in Parameter I because
$T_{\text{d}}$ is too low in this case. Then, can we expect the Chiral
Dominance in Parameter II? If so, it would be an ideal realization of
Satz's argument. The answer is, however, rather profound in fact.

For $m_q=1\,\text{MeV}$ with Parameter II, there appears only one peak
around $T_{\text{c}}\simeq 225\,\text{MeV}$. This peak results from
the second order chiral phase transition, which agrees with the thin
curve in Fig.~\ref{fig:order}. Thus the coincidence of the
pseudo-critical temperatures in this case can be understood according
to Satz's argument. Nevertheless this is not the Chiral Dominance
because of the theoretical requirement that chiral restoration occurs
at higher temperature than color-deconfinement does. We can prove that
the Gocksch-Ogilvie model satisfies this requirement \cite{fuk03}. We
can say that chiral restoration would be blocked by the Polyakov loop
behavior. Therefore, the simultaneous crossovers are really
\textit{simultaneous}. In other words, we have only one soft-mode
associated with the susceptibility peaks \cite{hat03}. We would
emphasize that this understanding is novel and non-trivial.

\begin{figure}[ht]
\begin{center}
\includegraphics[width=5cm]{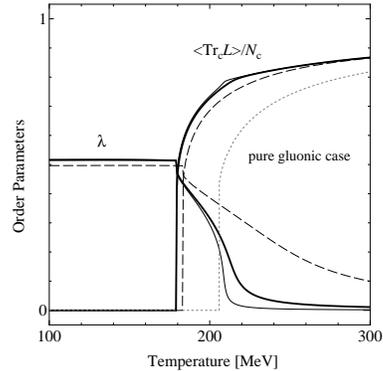}
\end{center}
\vspace{-10mm}
\caption{Order parameters for adjoint quarks with Parameter I.}
\label{fig:adjoint}
\vspace{-3mm}
\end{figure}

Finally, we shall present the results for adjoint quarks
($r=\text{adj}$) with Parameter I in Fig.~\ref{fig:adjoint}. In this
case, the center symmetry is not broken explicitly. The results
certainly show a clear first order transition in the Polyakov loop and
agree qualitatively with the lattice aQCD results \cite{kar99}.

In summary, the Gocksch-Ogilvie model can give a nice guideline to get
a deeper insight into the relation between the Polyakov loop dynamics
and the chiral dynamics. The underlying physics of simultaneous
crossovers is to be investigated more seriously in the future.

\end{document}